\documentclass[journal]{IEEEtran}
\usepackage{algorithmic}
\usepackage{color}
\usepackage{amsmath}
\usepackage{amsfonts}
\usepackage[pdftex]{graphicx}
\graphicspath{{fig/}}
\usepackage{threeparttable}
\usepackage{multirow}
\usepackage{makecell}
\usepackage{booktabs}
\usepackage[linesnumbered, ruled, vlined]{algorithm2e}
\usepackage{cite}
\usepackage{hyperref}
\usepackage{enumitem}

\begin{document}

\title{Temperature-Controlled Smart Charging for Electric Vehicles in Cold Climates
    \thanks{Manuscript accepted by IEEE Transactions on Smart Grid.}
    \thanks{G. Ruan and M. A. Dahleh are with the Laboratory of Information and Decision Systems, Massachusetts Institute of Technology, MA 02139, U.S.}
}

\author{
    Grant Ruan, 
    and Munther A. Dahleh
}

\maketitle

\begin{abstract}
The battery performance and lifespan of electric vehicles~(EVs) degrade significantly in cold climates, requiring a considerable amount of energy to heat up the EV batteries. 
This paper proposes a novel technology, namely temperature-controlled smart charging, to coordinate the heating/charging power and reduce the total energy use of a solar-powered EV charging station.
Instead of fixing the battery temperature setpoints, we analyze the thermal dynamics and inertia of EV batteries, and decide the optimal timing and proper amount of energy allocated for heating.
In addition, a temperature-sensitive charging model is formulated with consideration of dynamic charging rates as well as battery health.
We further tailor acceleration algorithms for large-scale EV charging, including the reduced-order dual decomposition and vehicle rescheduling.
Simulation results demonstrate that the proposed temperature-controlled smart charging is superior in capturing the flexibility value of EV batteries and making full use of the rooftop solar energy. The proposed model typically achieves a 12.5--18.4\% reduction in the charging cost and a 0.4--6.8\% drop in the overhead energy use for heating.
\end{abstract} 


\section{Introduction} \label{sec:intro}

\subsection{Background}

\begin{figure}[t]
    \centering
    \includegraphics[width=0.5\textwidth]{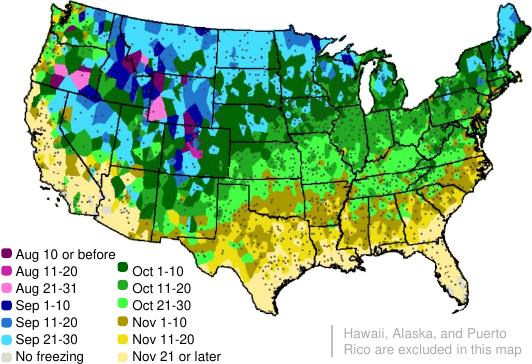}
    \caption{The first dates of a freezing median temperature across the U.S. mainland. The warm southern and coast regions may also experience sub-freezing temperature in November. The source data and graphic visualizations come from the U.S. national weather service.}
    \label{fig:stat}
\end{figure}

\IEEEPARstart{C}{old} climates and extreme weather are growing as a global concern for energy systems. Apart from high-latitude lands (with consistent cold weather or long winters), mid- or even low-latitude regions are also susceptible to harsh winter storms due to the disturbed climate dynamics~\cite{hook2021how}. Extreme cold is a popular statistical indicator for cold climates, which typically refer to a prolonged period of excessively cold weather (i.e., any temperature below freezing). As suggested in Fig.~\ref{fig:stat}, extreme cold is generally observed in all the U.S. states~\cite{noaa2022frost}. In Central New York, 132 days (36.2\% of the whole year) with sub-freezing temperature were recorded during 2021--2022, and 35 days were even all-day freezing~\cite{central2023freeze}. It might be surprising that the Plano city of Texas (the third hottest state) experienced 32 days (8.8\% of whole year) with sub-freezing temperature during 2021--2022~\cite{plano2023freeze}.

Cold climates, particularly extreme cold, are destructive to any battery packs of electric vehicles (EVs). A well-known bottleneck of lithium-ion batteries is that their performance and lifespan are relatively sensitive to temperature, while the low temperature is one of the most common causes of destruction in practice~\cite{liu2022perspective}. Here, the harms come from two aspects:
1)~Lower efficiency and longer time for charging: This is attributed to the large internal resistance in cold temperature~\cite{jaguemont2016comprehensive} as an electrolyte-level result of the higher solution conductivity and lower ion mobility.
2)~Reduced driving range and degraded battery capacity: This is due to the extra energy consumption from thermal management systems~\cite{zhang2018electric} for temperature control.
These two observations highlight the need to heat up EV batteries in cold climates. 

From a smart grid perspective, a practical question to ask is how these extra heating may shape the charging demand flexibility. The next follow-up question is how to coordinate heating and charging to achieve system-level benefits, such as plug-in EV preconditioning~\cite{wang2022low}.

In this paper, we propose a novel temperature-controlled smart charging technology to coordinate the EV charging and heating demand within a solar-powered charging station. The key idea is to put battery heating in the loop of smart charging so as to gain the system-level benefits of cost savings and climate resilience.

\subsection{Literature Review}

An extensive portion of the existing literature has studied the coordinated control and charging of EVs in a range of use cases.
Reference~\cite{wang2016smart} reported three types of EV charging: the grid-oriented, aggregator-oriented, and customer-oriented charging. Different kinds of coordination, such as \cite{zhang2021distributed,hu2013coordinated}, were carried out for each type, and the main idea was to explore the flexibility potential~\cite{ruan2021estimating} for financial revenue or other systematic benefits. The well-known smart charging and vehicle-to-grid services~\cite{turker2018optimal} have been studied on a rich set of topics such as control strategies, battery dynamics, driving patterns~\cite{hu2016electric}, and electric grid impacts~\cite{li2023centralized}.

Our work differs from the above efforts because we concentrate on the system-level coordination between charging (electric energy) and heating (thermal energy). The charging-related and heating-related research have been well developed in parallel, while very limited existing works were attempting to capture their strong couplings in cold climates.

There are a number of research efforts regarding the thermal management of EV batteries. Reference~\cite{kim2019review} summarized the major heat generation phenomena and thermal management issues for EV batteries, but the discussion was about cooling strategies. Reference~\cite{hamut2014analysis} introduced and analyzed the exergetic, exergoeconomic, and exergoenvironmental objectives in thermal management. Waste heat recovery was developed in~\cite{tian2020performance} to increase the coefficient of performance and driving mileage. A similar idea was applied in \cite{lee2013performance} for mobile heat pumps as well. Heat pumps~\cite{rao2013experimental} and heat pipes~\cite{kleiner2021influence} were two practical options for many applications in the literature. Understanding temperature impacts was important for thermal management. The authors from~\cite{zhang2022review} worked on the cold-induced security issues because lithium dendrites could appear in EV batteries in a freezing temperature. In~\cite{shelly2021comparative}, a detailed comparison of thermal management strategies was made across different temperature settings (-20--40{\textdegree}C) and different architectures, which suggested that waste heat recovery could empirically increase the vehicle mileage by 15\%.

Another research hotspot is the EV battery charging control in cold weather. Reference~\cite{bandhauer2014temperature} measured the heat effects during charging and formulated a functional mapping with the current and temperature inputs. Reference~\cite{liu2014analysis} broke down the heat generation during charging process into Joule heat and reaction heat. Then in \cite{ziat2021experimental}, the authors conducted experimental tests on charging-induced heat. They discovered a temperature drop at low state-of-charge and a constant temperature at high state-of-charge. This was an evidence that this kind of heat generation was physically nonlinear and state-dependent. Preconditioning an EV battery was mandatory in extreme cold~\cite{song2012experimental}, because the EV charging could only take place within a proper temperature range~\cite{chen2013design}. Reference~\cite{motoaki2018empirical} surveyed the EVs from New York and evaluated the severe impacts of cold temperature on EV performance. A well-known battery problem was that lithium-ion battery packs degraded rapidly in a cold winter~\cite{peng2019review}, but their performance could be recovered after being warmed up (8\% longer vehicle mileage and 3-6\% larger state-of-charge)~\cite{lindgren2016effect}
Note that the research works in this area often take the heating operation as a preparation step before charging, so there is no system-level coordination.

To summarize, there is limited knowledge on the optimal coordination strategy for EV charging and heating in cold climates. The existing efforts fail to develop a high-quality scheme to avoid potential degradation of limited flexibility, low efficiencies, and high overhead costs.

\subsection{Contributions}

We have made the following contributions in this paper:
\begin{enumerate}
\item Temperature-controlled smart charging is proposed for the first time to adapt EV smart charging to cold climates. The main idea is to integrate the thermal management into the energy management workflow and coordinate the heating and charging power at the system level. In this context, the charging demand flexibility can be accurately captured and utilized for system balancing.

\item A battery temperature control model is formulated to characterize the thermal transfer dynamics and variable heating power limits. Heat dissipation, heat transfer, and thermal inertia factors are taken into account.

\item A temperature-sensitive charging model is formulated to describe the low-temperature impacts. We follow a practical rule to mitigate capacity degradation by reducing charging rate in low temperature.

\item A reduced-order dual decomposition algorithm is developed for computational acceleration, and the vehicle rescheduling is applied to meet all system-wide constraints and achieve a reasonable level of solution quality in the case study.
\end{enumerate}

The reminder of this paper is organized as follows. Section~\ref{sec:system} presents the system framework and modeling setup for an EV charging station. The key modeling details of heating and charging dynamics are demonstrated in Section~\ref{sec:heat} and Section~\ref{sec:charge} accordingly. Computational algorithms are developed and discussed in Section~\ref{sec:accel}. Section~\ref{sec:case} conducts several numerical studies for validation, followed by the concluding remarks from Section~\ref{sec:concl}.

\section{System Framework} \label{sec:system}

\subsection{The Charging Station and Services}

\begin{figure}[t]
    \centering
    \includegraphics[width=0.48\textwidth]{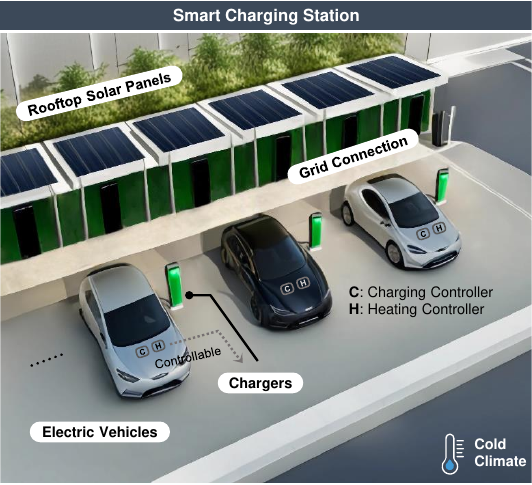}
    \caption{An illustrative smart charging station that supports charging and thermal management coordinately as a local microgrid. The rooftop solar panels and grid connection are available in this station system as well.}
    \label{fig:station}
\end{figure}

This paper is focused on the day-ahead scheduling of a grid-connected charging station in cold climates. This station serves as a local microgrid with EV chargers (Level-2 slow charging, AC power), rooftop solar panels, and grid connections. We assume that the station operator is able to communicate with each EV in a one-to-all and bidirectional communication network. This station is versatile to provide battery charging and thermal management services in a coordinated way to increase the demand flexibility and minimize the energy cost. Fig.~\ref{fig:station} visualizes a charging station of this kind in detail. 

Attention is given to the cold weather, and we assume that both the on-board chargers and heaters inside an EV are following an external signal from the charging station side. This allows a potential synergy between the charging and heating dynamics that differs from the common setup of local thermal management. 

The charging-as-a-service paradigm requires a charging station to offer not just charging energy but also maintenance supports such as temperature control. This innovative paradigm has great potential to unlock the system-level coordination and reduce the overall expense on EV management. Potential users include commuting service companies, battery swapping companies, and some customized charging stations. All of these stakeholders could expand their business and market share by attracting more EV users.

\subsection{Couplings Between Charging and Heating}

As mentioned in Section~\ref{sec:intro}, EVs face severe degradation in charging performances and battery service lifespan, and they may suffer more in freezing temperature. Therefore, thermal management is critical to keep EV batteries warm and ready for charging.

\begin{figure}[t]
    \centering
    \includegraphics[width=0.48\textwidth]{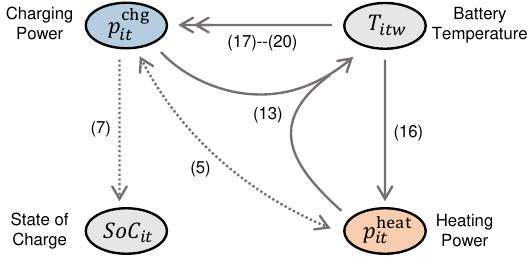}
    \caption{Illustration of the complicated couplings between four key decision variables from the thermal and energy management. All coupling arrows are marked with the associated formula (constraint) label in the main text.}
    \label{fig:coupling}
\end{figure}

The potential couplings of interest are demonstrated by four key variables: the charging power $p^\text{chg}_{it}$, the heating power $p^\text{heat}_{it}$, the battery temperature $T_{itw}$, and the state of charge~(SoC) $\mathit{SoC}_{it}$. Fig.~\ref{fig:coupling} uses arrows to express all the potential couplings and influential factors. We will explain the dotted arrows in the next subsection, and demonstrate the other couplings later in Section~\ref{sec:heat}~(solid arrows) and Section~\ref{sec:charge}~(a two-headed solid arrow). 

A good way to understand the couplings is through the scope of overhead energy use. Formally, overhead energy use refers to the extra electricity consumption that is utilized by a thermal management system to heat up the EV batteries. The couplings actually reflect a trade-off between the charging energy and the overhead energy use for keeping batteries warm. They could be technically expressed as several equality or inequality constraints using the four key variables shown in Fig.~\ref{fig:coupling}. 

The charging station seeks to capture these coupling factors and make an optimal trade-off to determine the best strategy for heating-charging coordination. This thereby empowers the cost-effective and sustainable EV charging.

\subsection{Temperature-Controlled Smart Charging}

In this subsection, we develop a novel technology called temperature-controlled smart charging, which is able to coordinate the EV charging and heating simultaneously. This technology is preferred to manage charging and heating within a unified workflow during cold weather.

In specific, the EV charging station makes a sub-hourly scheme (15 minutes) for the next day during 7am--10pm. Both the EV batteries and the on-board heaters are assumed to be fully controllable. In this station, the solar outputs and ambient temperature are two uncertainty sources modeled by scenarios. We assume relatively low (or roughly negligible) operational cost of solar panels, and we don't consider the vehicle queuing within the station.

A day-ahead scheduling model is formulated as follows:
\begin{align}
\min \quad 
& \sum\nolimits_t \sum\nolimits_w \pi_w \lambda_t \, p^\text{grid}_{tw} \Delta t \label{subeqn:obj} \\
\text{s.t.} \quad
& p^\text{pv}_{tw} + p^\text{grid}_{tw} = \sum\nolimits_i (p^\text{chg}_{it} + p^\text{heat}_{it}), \; \forall t, \forall w \label{subeqn:sys-balance} \\
& p^\text{grid}_{tw} \le \overline{pg}, \; \forall t, \forall w \label{subeqn:ub-pgrid} \\
& p^\text{pv}_{tw} \le \overline{pv}_{tw}, \; \forall t, \forall w \label{subeqn:ub-pv} \\
& p^\text{chg}_{it} + p^\text{heat}_{it} \le \overline{p}_i, \; \forall i, \forall t \in \Omega_i \label{subeqn:ub-pi} \\
& p^\text{chg}_{it} = p^\text{heat}_{it} = 0, \; \forall i, \forall t \notin \Omega_i \label{subeqn:set0} \\
& \mathit{SoC}_{i,t+1} = \mathit{SoC}_{it} + \eta_\text{chg} \, p^\text{chg}_{it} \Delta t / E_i, \; \forall i, \forall t \in \Omega_i \label{subeqn:soc-update} \\
& \mathit{SoC}_{i,\mathit{ta}_i} = \mathit{SoC}^\text{arr}_i, \; \forall i \label{subeqn:socarr} \\
& \mathit{SoC}_{i,\mathit{td}_i} \ge \mathit{SoC}^\text{dep}, \; \forall i \label{subeqn:socdep} \\
& 0 \le \mathit{SoC}_{it} \le 1, \; \forall i, \forall t \in \Omega_i \label{subeqn:soc-bnd} \\
& ( T_{itw}, p^\text{chg}_{it}, p^\text{heat}_{it} ) \in \Omega_\text{heat} \label{subeqn:f1} \\
& ( T_{itw}, p^\text{chg}_{it} ) \in \Omega_\text{chg} \label{subeqn:f2}
\end{align}
where $p^\text{pv}_{tw}, p^\text{grid}_{tw}, p^\text{chg}_{it}, p^\text{heat}_{it} \ge 0$ are the solar panel output, grid-side power injection, charging and heating power respectively; As previously mentioned, $\mathit{SoC}_{it}$ is the state of charge and $T_{itw}$ is the battery temperature; $\pi_w$ is the scenario probability; $\lambda_t$ is the time-of-use price; $\Delta t$ is 1/4 hour; $\overline{pv}_{tw}$, $\overline{pg}$, and $\overline{p}_i$ are three upper bounds; $\eta_\text{chg}$ is the charging efficiency; $E_i$ is the battery capacity; $\mathit{SoC}^\text{arr}_i$ and $\mathit{SoC}^\text{dep}$ are the SoC readings upon arrival and departure; $\Omega_i = [ \mathit{ta}_i, \mathit{td}_i ]$ covers the parking time steps of vehicle~$i$; $\Omega_\text{heat}$ and $\Omega_\text{chg}$ denote two prototype constraint sets to instantiate later.

The objective function~(\ref{subeqn:obj}) of the above model is used to minimize the expected operational cost, while the constraints include the system power balance~(\ref{subeqn:sys-balance}), the upper \& lower bounds for different decision variables~(\ref{subeqn:ub-pgrid})--(\ref{subeqn:set0})(\ref{subeqn:soc-bnd}), the SoC dynamics~(\ref{subeqn:soc-update}), and the required SoC status upon arrival and departure~(\ref{subeqn:socarr})(\ref{subeqn:socdep}). There are two additional prototype constraints~(\ref{subeqn:f1}) and (\ref{subeqn:f2}) being expressed in abstract forms. They refer to a group of constraints regarding the temperature control and temperature-sensitive charging process, and we will instantiate them with specific formulations later in the subsequent sections.

Furthermore, the above scheduling model applies an adaptive robust paradigm: The charging power $p^\text{chg}_{it}$ and heating power $p^\text{heat}_{it}$ are the ``here-and-now'' decisions made before observing the uncertain solar outputs and ambient temperature, while the power injection from the electric grid $p^\text{grid}_{tw}$ is a ``wait-and-see'' decision made after the realization of uncertainty factors.

\section{Battery Temperature Control} \label{sec:heat}

This section aims to model the thermal management process of EV batteries, including thermal transfer, thermal dynamics, and heating power limits. The models in this section will demonstrate the solid arrows in Fig.~\ref{fig:coupling} and instantiate the prototype constraint~(\ref{subeqn:f1}) accordingly.

\subsection{Heat Transfer and Dynamics}

The thermal dynamics of EV batteries are taking account of heat dissipation, heat transfer, and waste heat from charging power. This can be mathematically expressed by the following equality constraint:
\begin{align}
\label{eqn:thm-balance}
m_i c ( T_{i,t+1,w} - T_{itw} ) / \Delta t = - \mu_\text{heat} h A_i ( T_{itw} - T^\text{amb}_{tw} ) \notag \\
+ \eta_\text{heat} \, p^\text{heat}_{it} + (1 - \eta_\text{chg}) \, p^\text{chg}_{it}, \; \forall i, \forall w, \forall t \in \Omega_i
\end{align}
where $m_i$ is the battery mass; $c$ and $h$ are the heat capacity and the transfer coefficient; $T^\text{amb}_{tw}$ is the ambient temperature (can be negative); $\mu_\text{heat}$ is the thermal insulation coefficient; $\eta_\text{heat}$ is the heating efficiency that is influenced by heat transfer loss and uneven heating. 

\begin{figure}[t]
	\centering
	\includegraphics[width=0.48\textwidth]{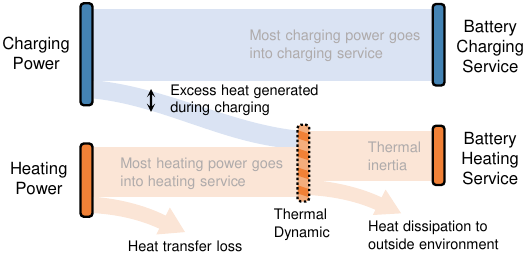}
	\caption{A Sankey diagram for electric and heating energy flows. Different energy transfer and transition processes are expressed as branch flows.}
	\label{fig:sankey}
\end{figure}

Fig.~\ref{fig:sankey} is a Sankey diagram visualizing the energy flows and energy balance conditions for EV batteries. It provides another way to express the thermal balance in~(\ref{eqn:thm-balance}).

The battery temperature should always be regulated within a preferred range. We assume the initial temperature of each EV battery is known (or self-reported) in advance. These settings are described in the following two constraints:
\begin{align}
& \underline{T} \le T_{itw} \le \overline{T}, \; \forall i, \forall w, \forall t \in \Omega_i \\
& T_{i,\mathit{ta}_i,w} = T^\text{arr}_i, \; \forall i, \forall w
\end{align}
where $\underline{T}$ and $\overline{T}$ are the lower and upper bounds of the preferred temperature range; $T^\text{arr}_i$ is the initial battery temperature. We typical regulate the battery temperature within 0--35{\textdegree}C.

\subsection{Heating Power Limits}

There are generally two options for the on-board heaters inside EVs: a positive temperature coefficient~(PTC) heater or a heat pump system. PTC heaters are often small-sized and perfect for fast and uniform heating; on the contrary, a heat pump system is large and efficient but may not be ideal for localized heating.

It holds true for both PTC heaters and heat pump systems that their heating efficiencies and capacities slightly vary across different temperature conditions~\cite{sorensen2023grid}. A well-known empirical observation is that the battery heating becomes more efficient when the temperature drops down~\cite{ziat2021experimental}.

This observation indicates a dynamic heating power limit, which can be linearly approximated as follows:
\begin{align}
p^\text{heat}_{it} \le \overline{ph}_i - \beta^\text{heat}_i \, T_{itw}, \; \forall i, \forall w, \forall t \in \Omega_i
\end{align}
where $\overline{ph}_i$ and $\beta^\text{heat}_i$ denote the regression intercept and slope coefficients to capture the variations.

\begin{figure}[t]
    \centering
    \includegraphics[width=0.48\textwidth]{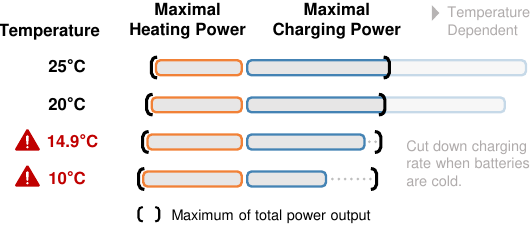}
    \caption{Heating and charging power limits under different temperature conditions. The temperature dependence is clear and the adverse consequences emerge as the temperature goes down below 15{\textdegree}C.}
    \label{fig:ub}
\end{figure}

Fig.~\ref{fig:ub} calculates the maximal heating power in different temperature. It well aligns with the above statement that heating limits slightly increase when the ambient temperature gets colder. 

\section{Temperature-Sensitive Charging} \label{sec:charge}

This section will illustrate how EV charging performance might be affected by cold temperature. A practical rule is followed to get rid of detrimental charging process. The models in this section will demonstrate the two-headed solid arrow in Fig.~\ref{fig:coupling} and instantiate the prototype constraint~(\ref{subeqn:f2}).

\subsection{Temperature Impacts on Charging Rate}

Temperature has an impact on charging rates and further shapes the performance of EV charging. Empirical studies such as~\cite{bandhauer2014temperature, motoaki2018empirical} have found a positive correlation between ambient temperature and charging performances. This finding extends the common assumption from smart charging literature and it suggests that the charging loss tend to increase in low temperature.

The temperature impact can be captured as follows:
\begin{align}
\label{eqn:chg-pwr-bnd}
p^\text{chg}_{it} \le \overline{pc}_i + \beta^\text{chg}_i \, T_{itw}, \; \forall i, \forall t, \forall w
\end{align}
where $\overline{pc}_i$ and $\beta^\text{chg}_i$ are the regression intercept and slope coefficients to approximate the potential correlation.

Fig.~\ref{fig:ub} shows the maximal charging power and one may observe a positive correlation between these maximum limits and temperature conditions.

\subsection{A Practical Rule to Avoid Severe Damage}

EV battery degrades heavily during high-current charging in low temperature. Practical evidences suggest that the degradation is proportional to the charging rate, which necessitates the low charging rates in cold temperature and no charging in sub-freezing temperature. 

Therefore, we apply a practical rule that greatly reduces the charging rate when the battery temperature drops below the setpoint. This rule is designed to avoid severe capacity degradation and it is formally expressed as follows:
\begin{align}
& p^\text{chg}_{it} \le \mu_\text{chg} T_{itw} + M (1 - v_{itw}), \; \forall i, \forall w, \forall t \in \Omega_i \label{subeqn:vitw1} \\
& v_{itw} \ge (T_\text{set} - T_{itw}) / M, \; \forall i, \forall w, \forall t \in \Omega_i \label{subeqn:vitw2}
\end{align}
where $v_{itw}$ is a binary variable detecting whether the battery temperature $T_{itw}$ is below the setpoint $T_\text{set}$. A cold temperature will lead to $v_{itw}=1$ (see (\ref{subeqn:vitw2})) and the constraint~(\ref{subeqn:vitw1}) is then activated to reinforce a much slower charging rate. See Fig.~\ref{fig:ub} for a graphic illustration.

Note that we do not assume specific models of capacity degradation due to their inconsistency (mostly case by case), nonlinearity (computational burden), and heterogeneity (hard for scheduling) in the literature. Since rule-based strategies are very common in practice, we decide to strictly follow the above practical rule, and in this case, the degradation level is guaranteed to be small enough.

\section{Computational Approach} \label{sec:accel}

This section will provide two computational approaches in either a centralized or a decentralized paradigm. The centralized approach is suitable for small-scale systems while the decentralized approach uses a heuristic to handle large-scale systems. Several acceleration tricks will be introduced and implemented.

\subsection{Centralized Approach}

\subsubsection{Binary Variable Reduction}

Reducing the number of binary variables can limit the search tree size in the branch-and-bound algorithm. We introduce another binary variable $v_{it}$ to denote the maximal value among all scenarios, i.e., $v_{it} = \max\nolimits_w v_{itw}, \forall i, \forall t$.

With this notation, we reformulate the previous constraint~(\ref{subeqn:vitw1}) and (\ref{subeqn:vitw2}) into the following way by substituting $v_{itw}$ with $v_{it}$. It is straightforward to prove the equivalence. 
\begin{align}
& p^\text{chg}_{it} \le \mu_\text{chg} T_{itw} + M (1 - v_{it}), \; \forall i, \forall w, \forall t \in \Omega_i \label{subeqn:vit1} \\
& v_{it} \ge (T_\text{set} - T_{itw}) / M, \; \forall i, \forall w, \forall t \in \Omega_i \label{subeqn:vit2}
\end{align}

After reformulation, the total number of binary variables is greatly reduced because the scenario index is removed in $v_{it}$.

\subsubsection{Overall Centralized Model and Solvers}

Taken together, the proposed centralized TCSC model is expressed as follows:
\begin{align}
\label{eqn:tcsc-cent}
\min \
& \text{(\ref{subeqn:obj})} \\
\text{s.t.} \
& \text{(\ref{subeqn:sys-balance})--(\ref{subeqn:soc-bnd}), (\ref{eqn:thm-balance})--(\ref{eqn:chg-pwr-bnd}), (\ref{subeqn:vit1})--(\ref{subeqn:vit2})} \notag
\end{align}
where the decision variables include $p^\text{pv}_{tw}$, $p^\text{grid}_{tw}$, $p^\text{chg}_{it}$, $p^\text{heat}_{it}$, $\mathit{SoC}_{it}$, $T_{itw}$, and $v_{it}$ (binary). Here, some rigid constraints such as (\ref{subeqn:socdep}) can be relaxed to soft constraints by adding additional slack variables and linear penalty terms.

The above model is a large-scale mixed-integer linear program~(MILP). Off-the-shelf optimization solvers are versatile to find the global optimum for MILP models by the branch-and-bound algorithm.

\subsection{Decentralized Approach}

\subsubsection{Reduced-Order Dual Decomposition}

It is clear that the proposed centralized model~(\ref{eqn:tcsc-cent}) has a partitioned constraint matrix, indicating that most constraints are vehicle-wise and separable. Such a special structure tends to align with the dual decomposition.

Now, the system-wide coupling from (\ref{subeqn:sys-balance}) can be used to remove $p^\text{grid}_{tw}$ in the centralized model~(\ref{eqn:tcsc-cent}) as follows:
\begin{align}
\min \; 
& \sum\nolimits_t \sum\nolimits_w \pi_w \lambda_t \left[ \sum\nolimits_i (p^\text{chg}_{it} + p^\text{heat}_{it}) - p^\text{pv}_{tw} \right] \Delta t \label{subeqn:dec-mdl-obj} \\
\text{s.t.} \; 
& \; 0 \le \sum\nolimits_i (p^\text{chg}_{it} + p^\text{heat}_{it}) - p^\text{pv}_{tw} \le \overline{pg}, \; \forall t, \forall w \label{subeqn:dec-mdl-coup-con} \\
& \text{(\ref{subeqn:ub-pv})--(\ref{subeqn:soc-bnd}), (\ref{eqn:thm-balance})--(\ref{eqn:chg-pwr-bnd}), (\ref{subeqn:vit1})--(\ref{subeqn:vit2})} \notag
\end{align}

We introduce a new variable $\hat{p}^\text{pv}_t = \min\nolimits_w p^\text{pv}_{tw}, \forall t$. Based on (\ref{subeqn:sys-balance}) and (\ref{subeqn:ub-pv}), it is direct to obtain $\hat{p}^\text{pv}_t$ by calculating the element-wise minimum as follows:
\begin{align}
\label{eqn:hat-ppv}
\hat{p}^\text{pv}_t = \min \! \left( \min\nolimits_w \overline{pv}_{tw}, \, \sum\nolimits_i (p^\text{chg}_{it} + p^\text{heat}_{it}) \right)
\end{align}

With this new variable, the constraint~(\ref{subeqn:dec-mdl-coup-con}) is equivalent to the following scenario-independent expression:
\begin{align}
\label{eqn:sim-con}
\sum\nolimits_i (p^\text{chg}_{it} + p^\text{heat}_{it}) - \hat{p}^\text{pv}_t \le \overline{pg}, \; \forall t
\end{align}

Clearly, the above reformulation (\ref{eqn:sim-con}) is more computational efficient then (\ref{subeqn:dec-mdl-coup-con}) because of a reduced number of constraints. It also leads to an order reduction of Lagrange multipliers and makes the iteration process easy to converge. 

We can design a proxy objective function that upper-bounds the original~(\ref{subeqn:dec-mdl-obj}) by replacing $p^\text{pv}_{tw}$ with $\hat{p}^\text{pv}_t$ below:
\begin{align}
\label{eqn:proxy-obj}
\min \; \sum\nolimits_t \lambda_t \left[ \sum\nolimits_i (p^\text{chg}_{it} + p^\text{heat}_{it}) - \hat{p}^\text{pv}_t \right] \Delta t
\end{align}

We next provide the formal expressions of the reduced-order dual decomposition model. Here, a Lagrange multiplier $\alpha_t$ is introduced to penalty the constraint~(\ref{eqn:sim-con}) as an additional term in the proxy objective function~(\ref{eqn:proxy-obj}). Technically, the decentralized model of vehicle~$i$ is given by:
\begin{align}
\label{eqn:indiv-model}
\min \; 
& \sum\nolimits_t (\lambda_t + \alpha_t) (p^\text{chg}_{it} + p^\text{heat}_{it}) \\
\text{s.t.} \; 
& \text{(\ref{subeqn:ub-pi})--(\ref{subeqn:soc-bnd}), (\ref{eqn:thm-balance})--(\ref{eqn:chg-pwr-bnd}), (\ref{subeqn:vit1})--(\ref{subeqn:vit2}) (remove $\forall i$)} \notag
\end{align}

Given $p^\text{chg}_{it}$, $p^\text{heat}_{it}$, and $\hat{p}^\text{pv}_t$ (updated by (\ref{eqn:hat-ppv})), the iteration formula for the Lagrange multiplier $\alpha_t$ is provided below:
\begin{align}
& \alpha_t \leftarrow \max \! \left(0, \, \alpha_t + \delta_t s \right), \; \forall t \label{eqn:lam-update} \\
& \delta_t = \sum\nolimits_i (p^\text{chg}_{it} + p^\text{heat}_{it}) - \hat{p}^\text{pv}_t - \overline{pg}, \; \forall t \label{eqn:delta-compute}
\end{align}
where $\delta_t$ is the excess power demand beyond the maximum limit; $s$ is a preset step size.

\subsubsection{Vehicle Rescheduling}

We select one (or some) balancing vehicle(s) to ensure the satisfaction of system-wide constraints by changing their own plans. Such a vehicle is often expected to be as flexible as possible, and we pick them by the following flexibility index:
\begin{align}
\label{eqn:flex-idx}
\mathit{FL}_i = \sum\nolimits_t \max (\delta_t, 0) \cdot ( p^\text{chg}_{it} + p^\text{heat}_{it} )
\end{align}

The above flexibility index $\mathit{FL}_i$ gives the highest values to the vehicle that can shift the greatest amount of power demand during congestion. The excess power $\delta_t$ acts as a weighting coefficient to prioritize the time with the heaviest congestion.

Let the vehicle~$i$ be the only balancing vehicle. An extra constraint for rescheduling can be easily derived from (\ref{eqn:sim-con}) by fixing the decisions of others:
\begin{align}
p^\text{chg}_{it} + p^\text{heat}_{it} \le \overline{pg} + \hat{p}^\text{pv}_t - \sum\nolimits_{l \neq i} (p^\text{chg}_{lt} + p^\text{heat}_{lt}), \; \forall t \label{eqn:new-con}
\end{align}

We formulate a rescheduling model from (\ref{eqn:indiv-model}) by dropping $\alpha_t$ in the objective function and supplementing the constraint (\ref{eqn:new-con}), shown as follows:
\begin{align}
\label{eqn:bal-model}
\min \; 
& \sum\nolimits_t \lambda_t (p^\text{chg}_{it} + p^\text{heat}_{it}) \\
\text{s.t.} \; 
& \text{(\ref{subeqn:ub-pi})--(\ref{subeqn:soc-bnd}), (\ref{eqn:thm-balance})--(\ref{eqn:chg-pwr-bnd}), (\ref{subeqn:vit1})--(\ref{subeqn:vit2}), (\ref{eqn:new-con}) (remove $\forall i$)} \notag
\end{align}

The above model can be easily extended to take two or more balancing vehicles into consideration. In a practical and loosely-congested case, the required number of balancing vehicles is often small.

\subsubsection{Overall Heuristic Approach}

Building upon the above models and index, the overall decentralized workflow can be summarized as follows:

\textbf{Step~1}: Initialize $\alpha_t=0, \forall t$, and prepare a group of step size options. For each option, run the vehicle-wise submodel~(\ref{eqn:indiv-model}), run the penalty update submodel~(\ref{eqn:lam-update}), and calculate the excess power demand $\delta_t$ using~(\ref{eqn:delta-compute}) for $N_\text{iter}$ times. If $\delta_t \le 0, \alpha_t = 0, \forall t$ holds true, the final solution is found, then output this solution and exit. After exploring all options, select the dual decomposition solution with the smallest $\max_t \delta_t$.

\textbf{Step~2}: Rank all the vehicles according to the flexibility index~(\ref{eqn:flex-idx}). Select the first vehicle that has the largest index value, that is $i^* = \arg\max_i \mathit{FL}_i$. Update the remaining excess power by $\delta_t \leftarrow \delta_t - p^\text{chg}_{i^*t} - p^\text{heat}_{i^*t}, \forall t$. If $\exists t, \delta_t > 0$, there is still some excess power after removing this vehicle. Then update the flexibility index~(\ref{eqn:flex-idx}) and select the next vehicle. This process is repeated until enough vehicles are chosen for system balancing.

\textbf{Step~3}: Reschedule the selected vehicles using~(\ref{eqn:bal-model}). Update the entire scheduling solution with the rescheduling results, and report the final decentralized solution.

\section{Case Study} \label{sec:case}

\subsection{Simulation Setup}

We conduct case studies in Boston during the last two months of 2022. As shown in Section~\ref{sec:system}, an EV charging station with rooftop solar panels and grid connections is designed to support the workplace charging during 7am--10pm for commercial buildings. 15-minute charging and heating power consumption are decided at the station level and on a day-ahead basis. In the subsections, we will first use 2--10 vehicles to demonstrate the benefits of temperature-controlled smart charging, and extend to consider 30--50 vehicles in the large-scale simulation.

We take into account the following four competing charging schemes, which are properly modified from the state-of-the-art approaches or industrial practices:
\begin{itemize}
\item \textbf{TCSC (Proposed)}: The proposed temperature-controlled smart charging scheme, categorized into TCSC-central (centralized approach) and TCSC-decent (decentralized approach). We set a computing time limit of 60~s/vehicle for TCSC-central and export the best solution ever found. As for TCSC-decent, the time limit is 10~s/vehicle for dual decomposition and 5~s/vehicle for rescheduling.

\item \textbf{SmartChg\&Heat}: The smart charging scheme with uncoordinated heaters. We run the classical smart charging model and reserve some capacity for thermal management. The entire workflow is to run smart charging and then run thermal management. The on-board heaters will keep the setpoint temperature during charging. We use a reservation ratio to keep capacity for heating power and this ratio is optimized between 15--30\%. Excess power beyond the allowable limit will be cut out at last.

\item \textbf{InstantChg\&Heat}: The instant charging scheme with uncoordinated heaters. On a first-come, first-served basis, this scheme starts charging once a vehicle is plugged in and it divides the available capacity equally for every vehicle. Localized thermal management, a reservation ratio between 15--30\%, and the excess power cut are all applied in the same way as SmartChg\&Heat.

\item \textbf{NoHeat}: The slow-speed smart charging scheme without heating. The charging rate in this scheme should be slow because of the practical rule for battery protection.
\end{itemize}

Simulation data are collected from several real-world data sources, and we perform data cleaning, harmonization, and integration to set up the case studies. 

In specific, the base capacity for EV batteries is 37~kWh, the base weight is 235.88~kg, and the base limit for total power output is 7.4~kW. We further introduce $\pm$5\% fluctuations in these base data to capture the vehicle-level differences. The arrival SoC is uniformly drawn from [0.0, 0.4], and the departure SoC is set as 0.9. The thermal management of EV batteries have a setpoint temperature of 15{\textdegree}C and a preferred temperature range of 0--35{\textdegree}C.

The parameters of charging-heating coupling are given as follows. The base values for $\overline{pc}_i$, $\beta_\text{chg}$, $\overline{ph}_i$, and $\beta_\text{heat}$ are 4.8, 0.12, 3.0, and 0.024. Additional $\pm$5\% fluctuations are added for each parameter. The efficiency coefficients include: $\eta_\text{chg}=0.92$, $\eta_\text{heat}=0.8$, $\mu_\text{chg}=0.22$, and $\mu_\text{heat}=0.4$. 

A time-of-use price signal is used to differentiate the on-peak (12pm--6pm, \textcent22.09/kWh), mid-peak (8am--12pm, \textcent17.22/kWh), and off-peak periods (before 8am or after 6pm, \textcent12.48/kWh). The daily average price \textcent17.50/kWh is comparable to the local retail prices in Boston. Note that the EV charging station purchases electricity from the grid at this time-of-use price (purchase price). 

Solar output data are obtained from real solar recordings in Boston (from ISO New England) and scaled down such that the maximal output is roughly 40\% or less of the total capacity. Two solar profiles from the target month are randomly selected, averaged, and fluctuated with $\pm$10\% noises to generate scenarios. We repeat this process to construct the scenario sets accordingly. Ambient temperature data are collected from an ASOS weather station in Boston, and the scenarios are generated by adding around $\pm$15\% fluctuations and $\pm$1{\textdegree}C shifts.

All the simulations are conducted on a laptop with Intel i7-8550U CPU and 16.0 GB RAM. The programming tool are Python 3.8.8, NumPy 1.20.1, Pandas 1.2.4, Pyomo 6.0.1, while the optimization solver is Mosek 9.2.

\subsection{Validation of Coordination Benefits}

The coordination benefits between charging and heating are first validated by an illustrative case of two vehicles and ten scenarios. The target day of interest is Nov. 21, 2022, with a potential temperature ranging from -3{\textdegree}C to 3.2{\textdegree}C.

Table~\ref{tab:case-stats} compares different charging schemes based on four statistical metrics: unmet SoC, charging cost, overhead energy use rate, and solar usage rate. All the metrics are calculated by averaging the results of different scenarios. In specific, the unmet SoC is the summation of all the final SoC differences below requirement (i.e. 0.9); the charging cost is defined as the total energy cost divided by the energy stored in EV batteries; the overhead energy use rate equals to the total heating energy divided by the total charging and heating energy consumption; and solar usage rate describes how much solar energy is utilized by the charging station. 

\begin{table}[t] 
    \caption{Statistics of different charging schemes for two-vehicle case}
    \label{tab:case-stats}
    \centering 
    \begin{threeparttable} 
        \begin{tabular}{p{7em} p{4.3em} p{4.3em} p{4.3em} p{4.3em}}
            \toprule
            Charging Scheme & Unmet SoC [p.u.] & Charging Cost [\textcent/kWh] & Overhead Rate [\%] & Solar Usage Rate [\%] \\
            \midrule
            TCSC-central     & 0.00 & 14.04 & \phantom{0}9.16 & 79.07 \\
            TCSC-decent      & 0.00 & 14.15 & \phantom{0}9.72 & 79.08 \\
            SmartChg\&Heat   & 0.04 & 16.31 & 16.22 & 75.99 \\
            InstantChg\&Heat & 0.00 & 15.86 & 11.02 & 78.41 \\
            NoHeat           & 1.06 & \phantom{0}5.92 & \phantom{0}0.00 & 37.85 \\
            \bottomrule
        \end{tabular}
    \end{threeparttable}
\end{table} 

Table~\ref{tab:case-stats} provides evidences to show the unique role of thermal management in cold temperature. In NoHeat, the absence of thermal management results in the slow charging rate and the final SoC deviates as much as 1.06 (or 70.3\%) in total. Using thermal management, all other charging schemes are able to reduce the unmet SoC, but there still exists a difference of 0.04 in SmartChg\&Heat. This is because SmartChg\&Heat (as well as InstantChg\&Heat) cannot always satisfy the system-wide constraints and the excessive demand will be cut out. 

Table~\ref{tab:case-stats} validates that the TCSC scheme is able to achieve the best cost savings and the greatest use of solar energy while satisfying all system-wide constraints and SoC requirements. Two TCSC approaches achieve an average cost of \textcent14.095/kWh, which is 12.5--15.7\% lower than SmartChg\&Heat and InstantChg\&Heat. This saving is partly due to the high photovoltaic utilization rate (0.7--3.1\% higher) and lower overhead energy use rate (1.6--6.8\% lower).

\begin{figure*}[t]
	\centering
	\includegraphics[width=0.98\textwidth]{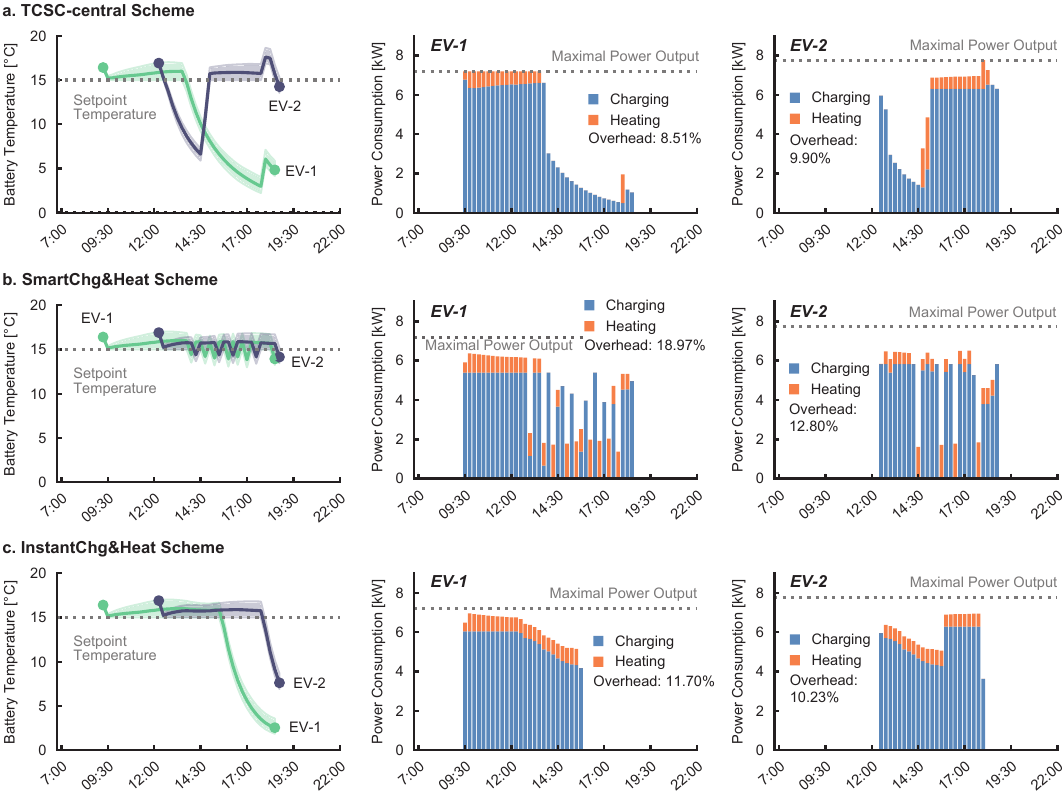}
	\caption{Operational details of different charging schemes for the two-vehicle case. The three rows denote the TCSC-central, SmartChg\&Heat, and InstantChg\&Heat schemes, while the first column shows the battery temperature dynamics with uncertainty bounds and the last two columns show the vehicle-level charging/heating demand separately. The overhead rates for each vehicle are given as well.}
	\label{fig:case-coord}
\end{figure*}

Fig.~\ref{fig:case-coord} presents the vehicle-level operating details to further demonstrate the cost savings of TCSC schemes. In the subfigures, we only simulate TCSC-central, SmartChg\&Heat, and InstantChg\&Heat for comparison.

Fig.~\ref{fig:case-coord} shows the TCSC's performance in the high-precision control of charging and heating dynamics. TCSC is able to make effective decisions on when and how much energy should be allocated for heating, and this is evidenced by its lowest overhead energy use rate for both vehicles (8.5\% and 9.9\%). In TCSC, the charging strategy for EV-1 (the first vehicle) is to steadily charge first and then relax the temperature tracking and switch into a slow charging rate. EV-2 shifts the charging demand for 2.5 hours to avoid system congestions and then raise the battery temperature to allow high charging rate afterwards. Such flexible switching and coordination is a unique feature of TCSC.

Fig.~\ref{fig:case-coord} also illustrates the potential cost burden of smart charging once it cannot align with localized heating. The thermal management strategy for SmartChg\&Heat and InstantChg\&Heat is to keep the battery temperature beyond the setpoint 15{\textdegree}C during EV charging. However, smart charging requires more heating power because the charging sequence is more dispersed, which cannot make full use of the thermal inertia. This finding is consistent with the statistical results in Table~\ref{tab:case-stats}, showing that the cheap charging of SmartChg\&Heat is offset by the surging demand of heating.

Table~\ref{tab:case-coord-more} runs another four groups of tests to verify the synergy for other cases. The test systems are the 5-vehicle 10-scenario case, the 5-vehicle 30-scenario case, the 10-vehicle 10-scenario case, and the 10-vehicle 30-scenario case.

\begin{table}[t] 
    \caption{Cost and overhead rate of different charging schemes under different scales of cases}
    \label{tab:case-coord-more}
    \centering
    \begin{threeparttable} 
        \begin{tabular}{p{3.5em} p{4.4em} p{4.4em} p{5.5em} p{5.5em}}
            \toprule
            Charging Scheme & Number of Vehicles & Number of Scenarios & Charging Cost {[}\textcent/kWh{]} & Overhead Rate {[}\%{]} \\
            \midrule
            \multirow[t]{4}{*}{\makecell[tl]{TCSC-\\central}} 
            & \phantom{0}5 & 10 & 12.87 & 10.25 \\
            & \phantom{0}5 & 30 & 13.72 & 10.69 \\
            & 10 & 10 & 15.12 & \phantom{0}9.62 \\
            & 10 & 30 & 14.96 & \phantom{0}9.74 \\
            \midrule
            \multirow[t]{4}{*}{\makecell[tl]{Smart-\\Chg\&Heat}} 
            & \phantom{0}5 & 10 & 15.94 & 17.47 \\
            & \phantom{0}5 & 30 & 16.59 & 17.27 \\
            & 10 & 10 & 17.37 & 14.98 \\
            & 10 & 30 & 17.19 & 15.09 \\
            \midrule
            \multirow[t]{4}{*}{\makecell[tl]{Instant-\\Chg\&Heat}} 
            & \phantom{0}5 & 10 & 16.15 & 10.27 \\
            & \phantom{0}5 & 30 & 16.88 & 10.35 \\
            & 10 & 10 & 16.90 & 10.59 \\
            & 10 & 30 & 16.75 & 10.83 \\
            \bottomrule
        \end{tabular}
    \end{threeparttable}
\end{table} 

Table~\ref{tab:case-coord-more} presents clear evidence that TCSC can achieve 17.7--18.4\% reduction in charging cost and 0.4--6.1\% drop in overhead energy use rate when compared to SmartChg\&Heat and InstantChg\&Heat. These results support the previous finding that TCSC is versatile to decide good allocation for charging and heating power such that the charging cost per kWh and overhead energy use rates are effectively minimized.

\subsection{Impacts of Ambient Temperature}

Ambient temperature has a significant impact on the efficiency and expense of charging schemes. Lower temperature generally leads to a stronger coupling between charging and heating and that calls for the use of TCSC schemes.

This subsection is focused on the 10-vehicle 10-scenario case on Nov. 21 with additional shifts of -9--9{\textdegree}C on top of the ambient temperature to represent different scenarios. We typically simulate TCSC-central, SmartChg\&Heat, and InstantChg\&Heat schemes for comparison.

\begin{figure}[t]
    \centering
    \includegraphics[width=0.48\textwidth]{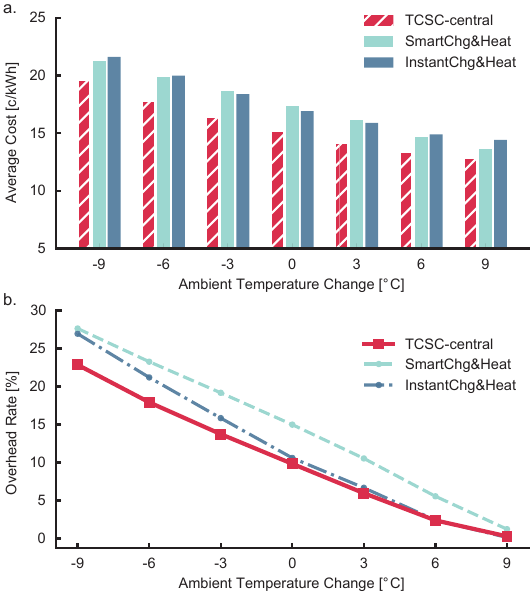}
    \caption{Temperature impacts comparison between different charging schemes. In a ten-vehicle case, diverse levels of changes are added to the ambient temperature of Nov. 21, 2022, and the associated charging costs and overhead energy use rates of TCSC-central, SmartChg\&Heat, and InstantChg\&Heat are given accordingly.}
    \label{fig:case-tmpc-day}
\end{figure}

Fig.~\ref{fig:case-tmpc-day} shows a clear trend that the charging costs and overhead energy use rates steadily increase as the ambient temperature is dropping down, and this finding holds true for all three schemes. In particular, the temperature impacts on TCSC is less than the other two. For TCSC, every temperature drop of 1{\textdegree}C may lead to \textcent0.37/kWh increase in the charging cost and 1.3\% increase in the overhead energy use rate. The sensitivities for SmartChg\&Heat and InstantChg\&Heat are estimated to be \textcent0.43/kWh, 1.5\%, \textcent0.41/kWh, 1.5\% accordingly. In other word, TCSC is effective to improve the climate resilience and adaptation by reducing the extra energy demand induced by cold temperature.

In Fig.~\ref{fig:case-tmpc-day}, the TCSC's benefits of cost savings and reduced overhead energy are enhanced in colder temperature. Taking overhead energy as an example, the gaps between TCSC and the other two schemes are increasing when temperature becomes lower. Furthermore, both SmartChg\&Heat and InstantChg\&Heat cannot meet the SoC requirement with negative temperature shifts, and the unmet SoC increases from 0.05, 0.02 for 3{\textdegree}C colder to 0.37, 0.25 for 9{\textdegree}C colder. But for TCSC, it satisfies all system-wide constraints and requirements for any temperature shifts. Taken together, TCSC is the only climate-resilient scheme that can secure the charging performance at the minimal charging expense.

\begin{figure}[t]
    \centering
    \includegraphics[width=0.48\textwidth]{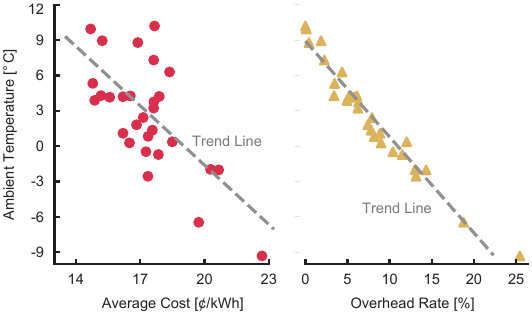}
    \caption{Temperature impact assessment for a ten-vehicle case in Dec. 2022. Daily simulation of TCSC-central scheme is operated for a whole month and the correlations between ambient temperature, average charging cost, and overhead energy use rate are demonstrated using two linear trend lines. The two subfigures share a y-axis for visual simplification.}
    \label{fig:case-tmpc-scan}
\end{figure}

We next simulate the entire month of December 2022 at a daily resolution to validate the robust performance of TCSC. The month-long results are collected and visualized in Fig.~\ref{fig:case-tmpc-scan}.

Fig.~\ref{fig:case-tmpc-scan} presents a similar observation of the negative correlation between ambient temperature (daily average) and average charging cost as well as overhead energy use rate. Different temperature profiles can influence the final charging cost and cause more fluctuations, but the entire negative correlation is still evident. Regarding the sensitivities, every 1{\textdegree}C temperature drop may lead to \textcent0.24/kWh cost growth and 1.2\% higher overhead energy use in December 2022. It might not be surprising that the correlation between ambient temperature and overhead rate is stronger than charging cost, indicating that overhead is largely influenced and driven by temperature-related factors.

\subsection{Large-Scale Implementation}

This subsection is focused on the large-scale simulation to compare the computational speed and optimum searching efficiency for TCSC-central and TCSC-decent.

We apply the tests on three selected days of December, i.e., Dec. 12--14. The 30-vehicle 60-scenario case, the 30-vehicle 100-scenario case, the 50-vehicle 60-scenario case, and the 50-vehicle 100-scenario case are evaluated in terms of running time and charging cost.

Table~\ref{tab:case-large-scale} collects all the results and statistics for comparison. It is validated that TCSC-decent is more computationally efficient than TCSC-central when dealing with large-scale problems. In particular, TCSC-decent runs faster for nearly all cases and the largest time saving is almost 30.0\% for the 30-vehicle 60-scenario case. Regarding the charging cost, TCSC-decent achieves slightly better performance than TCSC-central in most cases. On average, TCSC-decent spends 21.3\% less time in 30-vehicle cases and 12.8\% less time in 50-vehicle cases to make better solutions with cost savings of \textcent0.18/kWh.

\begin{table}[t] 
    \caption{Large-scale tests for two TCSC computational solutions.}
    \label{tab:case-large-scale}
    \centering
    \begin{threeparttable} 
        \begin{tabular}{p{1.8em} p{2.8em} p{2.8em} p{3em} p{3em} p{3em} p{3em}}
            \toprule
            Date & \multirow[t]{2}{*}{\makecell[tl]{Number\\of Vehi-\\cles}} & \multirow[t]{2}{*}{\makecell[tl]{Number\\of Scen-\\arios}} & \multicolumn{2}{l}{TCSC-central} & \multicolumn{2}{l}{TCSC-decent} \\
            \cmidrule(lr{1em}){4-5} \cmidrule(lr{1em}){6-7}
            &  &  & Time [min] & Cost [\textcent/kWh] & Time [min] & Cost [\textcent/kWh] \\
            \midrule
            \multirow[t]{4}{*}{\makecell[tl]{Dec.\\12}} 
            & 30 & \phantom{0}60 & 30.69 & 20.47 & 24.81 & 20.26 \\
            & 30 & 100 & 30.89 & 20.31 & 28.72 & 20.32 \\
            & 50 & \phantom{0}60 & 51.12 & 21.15 & 42.58 & 20.76 \\
            & 50 & 100 & 51.55 & 20.87 & 52.38 & 20.80 \\
            \midrule
            \multirow[t]{4}{*}{\makecell[tl]{Dec.\\13}} 
            & 30 & \phantom{0}60 & 30.59 & 17.64 & 18.72 & 17.33 \\
            & 30 & 100 & 31.15 & 17.47 & 24.53 & 17.25 \\
            & 50 & \phantom{0}60 & 51.34 & 18.17 & 37.63 & 17.91 \\
            & 50 & 100 & 51.61 & 17.94 & 48.54 & 17.86 \\
            \midrule
            \multirow[t]{4}{*}{\makecell[tl]{Dec.\\14}} 
            & 30 & \phantom{0}60 & 30.63 & 17.75 & 21.46 & 17.52 \\
            & 30 & 100 & 31.13 & 17.63 & 27.42 & 17.53 \\
            & 50 & \phantom{0}60 & 51.28 & 18.14 & 42.88 & 18.02 \\
            & 50 & 100 & 51.86 & 18.04 & 45.27 & 17.87 \\
            \midrule
            \multicolumn{3}{l}{Average for 30 Vehicles} & 30.85 & 18.55 & 24.28 & 18.37 \\
            \multicolumn{3}{l}{Average for 50 Vehicles} & 51.46 & 19.05 & 44.88 & 18.87 \\
            \bottomrule
        \end{tabular}
    \end{threeparttable}
\end{table} 

Note that both TCSC-central and TCSC-decent are terminated after reaching their time limits in Table~\ref{tab:case-large-scale}. It is observed that the centralized approach is less efficient and the relative gaps may converge very slowly after shrinking below 10\%. On the contrary, the decentralized approach implements a group of small-scale submodels that are more effective in solution seeking within limited time.

In addition, Table~\ref{tab:case-large-scale} does not consider the space complexity. TCSC-central appears to have a higher requirement for RAM memory than TCSC-decent, so the memory shortage may occur in devices with limited computing resources. Another feature of TCSC-decent is the potential to implement acceleration by taking advantage of parallel computing. For instance, the four-core parallel computing can roughly accelerate the total running time by a factor of four.

We can finally summarize that TCSC-decent is more preferred for large-scale systems due to its high searching efficiency, and it could be further accelerated using parallel computing; TCSC-central is suitable to search (nearly) global optimum for small- to moderate-scale systems.

\section{Conclusion} \label{sec:concl}

In cold climates, the performance and health of EV batteries severely degrade. A common remedy is to heat these batteries and conduct thermal management. Such extra heating may greatly limit the demand-side flexibility, and ignoring the potential charging-heating coordination could harm the system performance and climate resilience.

This paper proposes a novel temperature-controlled smart charging technology to boost the cost effectiveness and climate adaptation of a solar-powered EV charging station in cold climates. This technology, with heating integrated into charging management, can unlock the demand-side flexibility by controlling the battery temperature and charging rates. 

Technically, we develop a battery temperature control model to analyze the battery thermal dynamics, and a temperature-sensitive model to charge EV batteries at a 15-minute resolution. Numerical results indicate that the proposed TCSC scheme could accurately explore the operational potential of EV batteries in cold climates, and achieve a 12.5--18.4\% reduction in energy cost across different model scales and weather conditions. 

The proposed concepts, models, approaches, and findings have opened up new opportunities for the future design of EV-integrated energy systems (e.g., charging stations, microgrids), especially those located in high-latitude countries. 

As for future work, extra efforts are needed to analyze the space-limited case with EV queuing~\cite{said2019novel},  explore the vehicle-to-grid potential to provide more electricity market services~\cite{ruan2022improving}, and implement reinforcement learning for fast EV scheduling~\cite{qiu2023reinforcement}. These future works are expected to extend the methodologies and findings of this paper to other use cases.

\appendices

\section{Nomenclature}

\begin{itemize}[leftmargin=4.5em,style=nextline]
    \item[$M, \epsilon$] A large and a small constant.
    \item[$i,t,w$] Indices of vehicles, time steps, and scenarios.
    
    \item[$p^\text{chg}_{it}, p^\text{heat}_{it}$] Charging and heating power of vehicle~$i$ and time~$t$.
    \item[$p^\text{pv}_{tw}$] Solar panel output of time~$t$ and scenario~$w$.
    \item[$p^\text{grid}_{tw}$] Grid-side power injection of time~$t$ and scenario~$w$.
    \item[$\mathit{SoC}_{it}$] State of charge of vehicle~$i$ and time~$t$.
    \item[$T_{itw}$] Battery temperature of vehicle~$i$, time~$t$, scenario~$w$.
    \item[$v_{itw}, v_{it}$] Binary variables indicating whether the battery temperature of a specific case is below the setpoint (value=1) or not (value=0). 
    
    \item[$\Delta t$] Duration between two time steps.
    \item[$c, h$] Specific heat capacity coefficient and heat transfer coefficient for EV batteries.
    \item[$\pi_w$] Probability of scenario~$w$.
    \item[$\lambda_t$] Time-of-use price at time~$t$.
    
    \item[$E_i, m_i$] Battery capacity and mass of vehicle~$i$.
    \item[$\mathit{SoC}^\text{arr}_i$] Initial state of charge of vehicle~$i$ upon arrival.
    \item[$\mathit{SoC}^\text{dep}$] Required level of state of charge before departure.
    \item[$T^\text{arr}_i$] Initial battery temperature of vehicle~$i$ upon arrival.
    \item[$T^\text{amb}_{tw}$] Ambient temperature at time~$t$ and in scenario~$w$.
    \item[$\mathit{ta}_i, \mathit{td}_i$] Arrival and departure time of vehicle~$i$.
    
    \item[$\overline{p}_i$] Maximum limits of total power output.
    \item[$\overline{pc}_i,\beta_\text{chg}$] Coefficients describing the temperature impact on the charging power limits.
    \item[$\overline{ph}_i,\beta_\text{heat}$] Coefficients describing the temperature impact on the charging heating limits.
    \item[$\overline{pg}$] Maximum limits of grid-side power injection.
    \item[$\overline{pv}_{tw}$] Maximal output scenarios of solar panels. 
    \item[$\hat{p}^\text{pv}_t$] Potentially minimal solar output at time~$t$.
    \item[$\underline{T}, \overline{T}$] Preferred temperature range for EV batteries.
    \item[$T_\text{set}$] Setpoint temperature for thermal management.
    
    \item[$\eta_\text{chg}, \eta_\text{heat}$] Charging and heating efficiency coefficient.
    \item[$\mu_\text{chg}$] Charging rate reduction due to cold climate. 
    \item[$\mu_\text{heat}$] Thermal insulation efficiency coefficient.
    
    \item[$\Omega_i$] Duration of vehicle~$i$ in the charging station.
    \item[$\Omega_\text{heat}, \Omega_\text{chg}$] Prototype constraint sets for battery temperature control and temperature-sensitive charging.
    
    \item[$\alpha_t$] Iterative step size in dual decomposition.
    \item[$\delta_t$] Excess power demand beyond the limits.
    \item[$\mathit{FL}_i$] Flexibility index for choosing balancing vehicles.
    \item[$N_\text{iter}$] Number of dual decomposition iterations.
\end{itemize}

\bibliographystyle{ieeetr}
\bibliography{refs}

\end{document}